\documentclass{article}

\usepackage{arxiv}

\usepackage[utf8]{inputenc} 
\usepackage[T1]{fontenc}    
\usepackage{hyperref}       
\usepackage{url}            
\usepackage{booktabs}       
\usepackage{amsfonts}       
\usepackage{nicefrac}       
\usepackage{microtype}      
\usepackage{lipsum}		
\usepackage{graphicx} 
\usepackage{amsmath}
\usepackage{gensymb}
\usepackage{amssymb}
\usepackage{setspace}  
\usepackage[english]{babel}
\usepackage[]{natbib}

\newcommand{\apj}{    {\it Astrophys. J.}}
\newcommand{\apjl}{   {\it Astrophys. J. Lett.}}

\newcommand{\jgr}{    {\it J. Geophys. Res.}}

\newcommand{\solphys} {{\it Solar Phys.}}

\title{Forecasting the Ambient Solar Wind with Numerical Models: I. On the Implementation of an Operational Framework}

\date{2018 December 12}	

\author{
  Martin A.~Reiss \\
  Heliophysics Science Division\\
  NASA Goddard\\
  Greenbelt, MD 20771, USA \\
  \texttt{martin.reiss@nasa.gov} \\
  \AND
  Peter J.~MacNeice \\
  Heliophysics Science Division\\
  NASA Goddard \\
  Greenbelt, MD 20771, USA \\
  \And
  Leila M.~Mays \\
  Heliophysics Science Division\\
  NASA Goddard \\
  Greenbelt, MD 20771, USA \\
  \And
  Charles N. Arge \\
  Heliophysics Science Division\\
  NASA Goddard\\
  Greenbelt, MD 20771, USA \\
  \And
  Christian M\"ostl \\
  Space Research Institute \\
  Austrian Academy of Sciences \\
  8042 Graz, Austria \\
  \And
  Ljubomir Nikolic \\
  Canadian Hazards Information Service \\
  Natural Resources Canada \\ 
  Ottawa, Canada \\
  \And
  Tanja Amerstorfer \\
  Space Research Institute \\
  Austrian Academy of Sciences \\
  8042 Graz, Austria \\
}

\begin{document}
\maketitle

\begin{abstract}
The ambient solar wind conditions in interplanetary space and in the near-Earth environment are determined by activity on the Sun. Steady solar wind streams modulate the propagation behavior of interplanetary coronal mass ejections and are themselves an important driver of recurrent geomagnetic storm activity. The knowledge of the ambient solar wind flows and fields is thus an essential component of successful space weather forecasting. Here, we present an implementation of an operational framework for operating, validating, and optimizing models of the ambient solar wind flow on the example of Carrington Rotation 2077. We reconstruct the global topology of the coronal magnetic field using the potential field source surface model (PFSS) and the Schatten current sheet model (SCS), and discuss three empirical relationships for specifying the solar wind conditions near the Sun, namely the Wang-Sheeley (WS) model, the distance from the coronal hole boundary (DCHB) model, and the Wang-Sheeley-Arge (WSA) model. By adding uncertainty in the latitude about the sub-Earth point, we select an ensemble of initial conditions and map the solutions to Earth by the Heliospheric Upwind eXtrapolation (HUX) model. We assess the forecasting performance from a continuous variable validation and find that the WSA model most accurately predicts the solar wind speed time series (RMSE $\approx 83$~km/s). We note that the process of ensemble forecasting slightly improves the forecasting performance of all solar wind models investigated. We conclude that the implemented framework is well suited for studying the relationship between coronal magnetic fields and the properties of the ambient solar wind flow in the near-Earth environment.
\end{abstract}


\section{Introduction} \label{sec:introduction}

The knowledge of the evolving ambient solar wind is an essential component of successful space weather forecasting~\citep{owens13}. The ambient solar wind flows and fields play an important role in understanding the propagation of coronal mass ejections (CMEs) and are themselves an important driver of recurrent geomagnetic activity. More specifically, high-speed solar wind streams contribute about $70\, \%$ of geomagnetic activity outside of solar maximum and about $30\, \%$ at solar maximum~\cite{richardson00}. Even when the occurrence rate of CMEs increases from $0.3$ per day during solar minimum to about $4$--$5$ per day during solar maximum~\cite{stcyr99}, key properties in interplanetary space, such as the solar wind bulk speed, magnetic field strength, and orientation, are determined by the ambient solar wind flow~\cite{luhmann02}.  

\begin{figure*}
\begin{center}
\includegraphics[width=0.99\columnwidth]{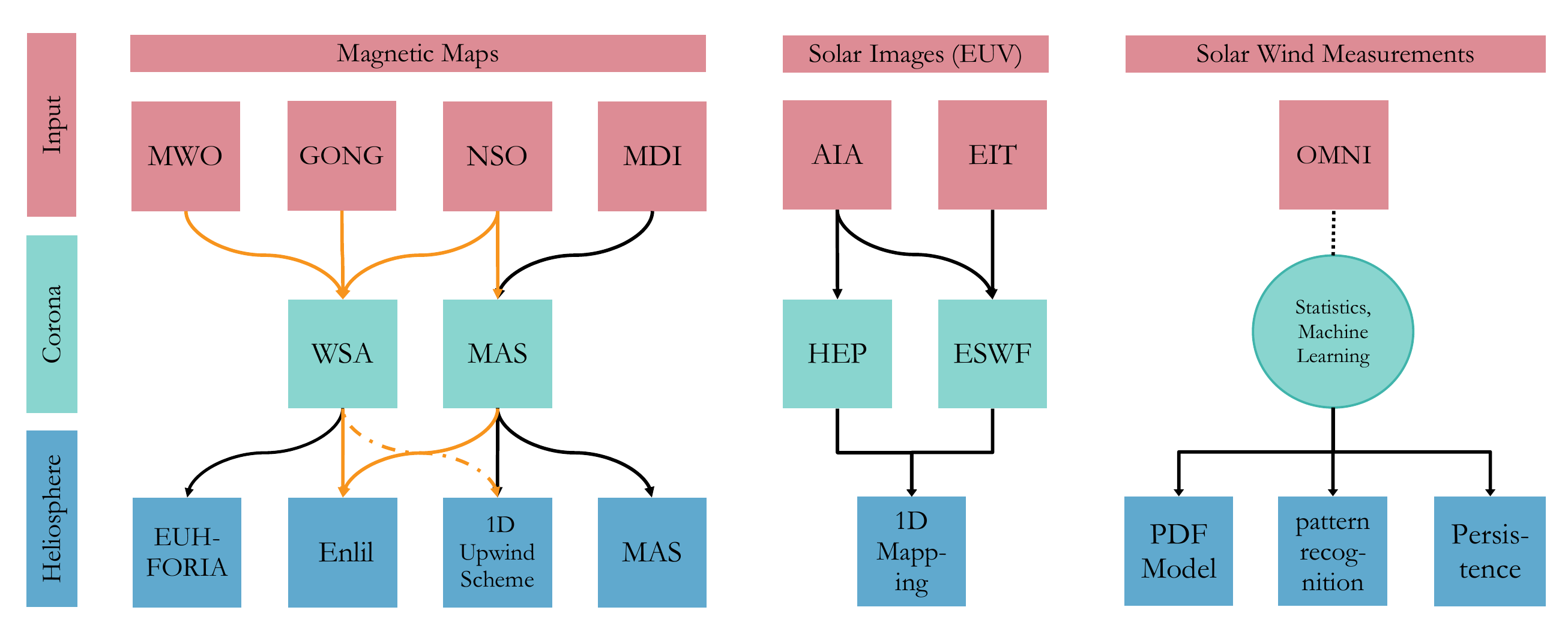}
\end{center}
\caption{Overview of coronal and heliospheric model combinations for forecasting the ambient solar wind conditions in the near-Earth environment. The orange colored lines highlight models hosted at NASA's Community Coordinated Modeling Center online platform (see, \url{https://ccmc.gsfc.nasa.gov/models/}). \label{fig:figure1}}
\end{figure*}

The topology of open magnetic field lines along which ambient solar wind flows accelerate to supersonic speeds plays a fundamental role in understanding phenomena that drive our evolving space weather. To date, there are no routine measurements of the coronal magnetic field. Magnetic models for the solar corona, therefore, have to rely on extrapolations calculated from the observed line-of-sight component of the photospheric field. The most widely applied extrapolation technique to reconstruct global solutions for the coronal magnetic field is the Potential Field Source Surface model~\cite[PFSS;][]{altschuler69,schatten69}. Using the current-free (or potential field) assumption for regions above the photosphere, solutions for the magnetic field can be expressed as the gradient of a magnetic scalar potential. Since potential fields give closed magnetic fields, a spherical source surface, where the magnetic field is assumed to be only radial is added as an outer boundary condition. The radius of the spherical source surface is an adjustable parameter typically set to a reference height of $2.5 \,R_{0}$ to best match observations~\cite{hoeksema82}. To account for \textit{Ulysses} measurements showing latitudinal invariance of the radial magnetic field component~\cite{wang95}, the so-called Schatten current sheet (SCS) model~\cite{schatten71} is typically added beyond the PFSS model to create a more uniform radial field strength solution. 

The effect of the solar wind is to drag and distort magnetic field lines, and thus to distort the coronal field from the assumed current-free configuration. Ideally, a model should account for the complex dynamics of the solar wind flow by solving a set of nonlinear partial differential equations of magnetohydrodynamics (MHD)~\cite[e.g.,][]{altschuler69}. As such, the Magnetohydrodynamics Algorithm outside a Sphere model~\cite[MAS;][]{linker99,mikic99} and the Space Weather Modeling Framework~\cite[SWMF;][]{toth05} are three-dimensional MHD models. The PFSS solutions are usually used as an initial condition, and the MHD equations are integrated in time until the plasma and magnetic fields relax into equilibrium. The final steady-state solution is characterized by closed magnetic fields that confine the solar wind plasma and open magnetic field lines along which solar wind flows accelerate to supersonic speeds.

The state-of-the-art framework for forecasting the ambient solar wind couples models of the corona with those of the inner heliosphere~\cite{riley01, lee08}. The coronal part spans the range from 1~solar radii ($R_{0}$) to $2.5 \,R_{0}$ (PFSS), or $20 \,R_{0}$ to $30 \,R_{0}$ (MHD). The inner boundary condition of the heliospheric part either results directly from the coronal model or is computed from the topology of the coronal magnetic field. The outer boundary condition computed from the coronal model is used as inner boundary condition for the heliospheric model ($20$--$30 \,R_{0}$ to 1~au). Typically, the heliospheric model is based on the MAS model~\cite{linker99, mikic99}, Enlil~\cite{odstrcil03}, or the recently developed European Heliospheric Forecasting Information Asset~\cite[EUHFORIA;][]{pomoell18}, each of which are three-dimensional numerical MHD models that derive stationary solutions for the ambient solar wind in interplanetary space. Figure~\ref{fig:figure1} shows some different coronal and heliospheric model combinations, together with other approaches for forecasting the ambient solar wind based on empirical relationships~\cite[e.g.,][]{robbins06, vrsnak07, shugay11, reiss16}, and statistics or machine learning techniques~\cite[e.g.,][]{owens13b,riley17}. The orange colored lines highlight models accessible online at NASA's Community Coordinated Modeling Center (CCMC) online platform.

\begin{figure*}
\begin{center}
\includegraphics[width=0.99\columnwidth]{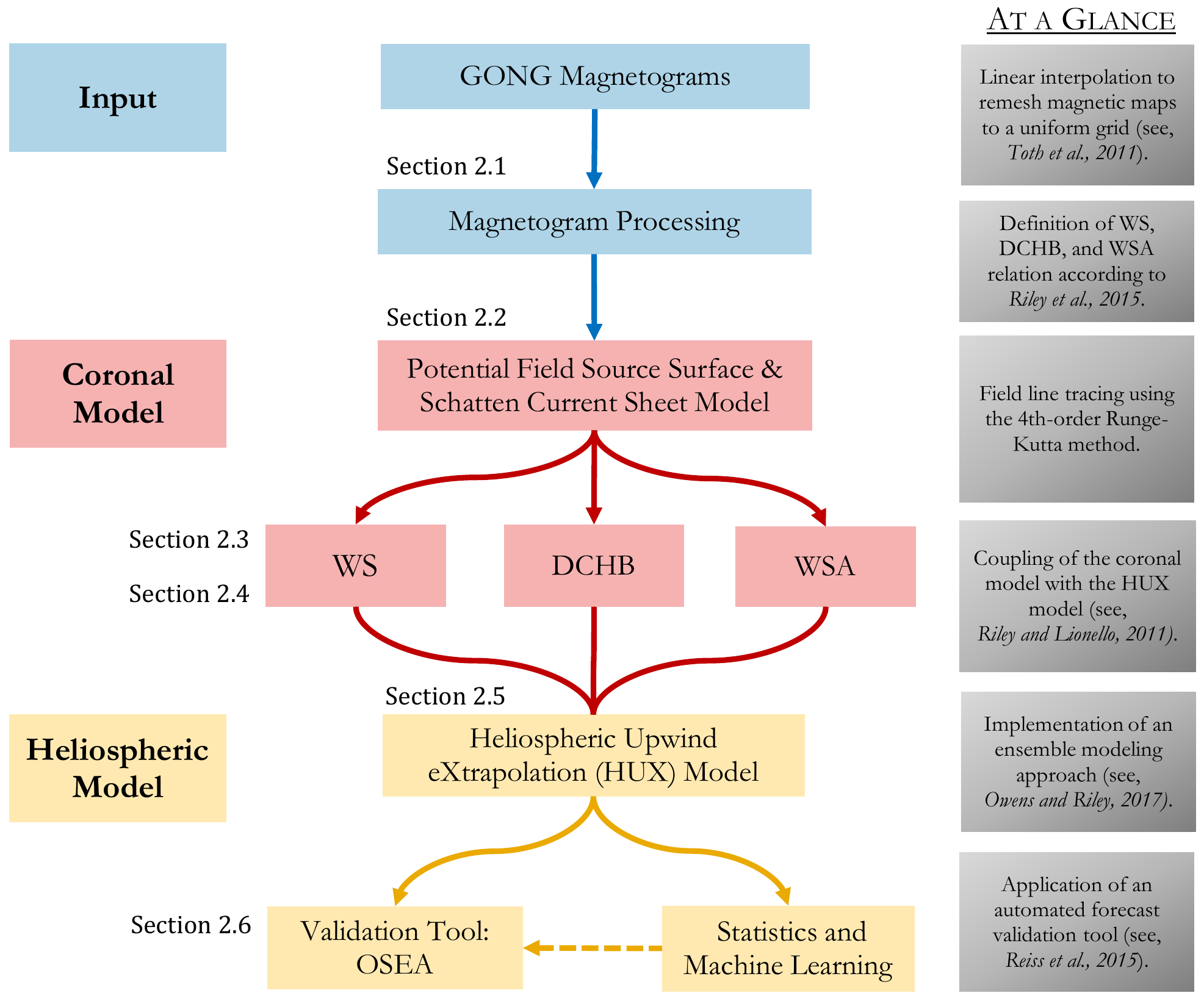}
\end{center}
\caption{Breakdown structure of the framework implementation. The sections explaining the corresponding framework component are indicated. \label{fig:figure2}}
\end{figure*}

Many recent studies~\cite[e.g.,][]{owens08,jian11,devos14,reiss16} have assessed the performance of operational frameworks for forecasting the ambient solar wind and reported typical uncertainties of about 1~day in the arrival time of high-speed streams. Furthermore, it is now well established that the performance of models of the ambient solar wind is, if at all, only slightly better than a 27 day persistence model~\cite[e.g.,][]{owens13b,kohutova16}, assuming that the near-Earth solar wind conditions will replicate after each Carrington Rotation (CR). Overall, these results highlight the fact that forecasting the conditions in the ambient solar wind in interplanetary space and in the near-Earth environment is challenging, even during times of low solar activity when the large-scale interplanetary magnetic field configuration evolves less rapidly and disturbances due to CMEs are less frequent~\cite{owens13}. As outlined in the space weather roadmap for the years 2015--2025~\cite[see,][]{schrijver15}, continued efforts are needed to improve our capabilities for forecasting the conditions in the ambient solar wind.

\begin{figure}
\begin{center}
\includegraphics[width=0.75\columnwidth]{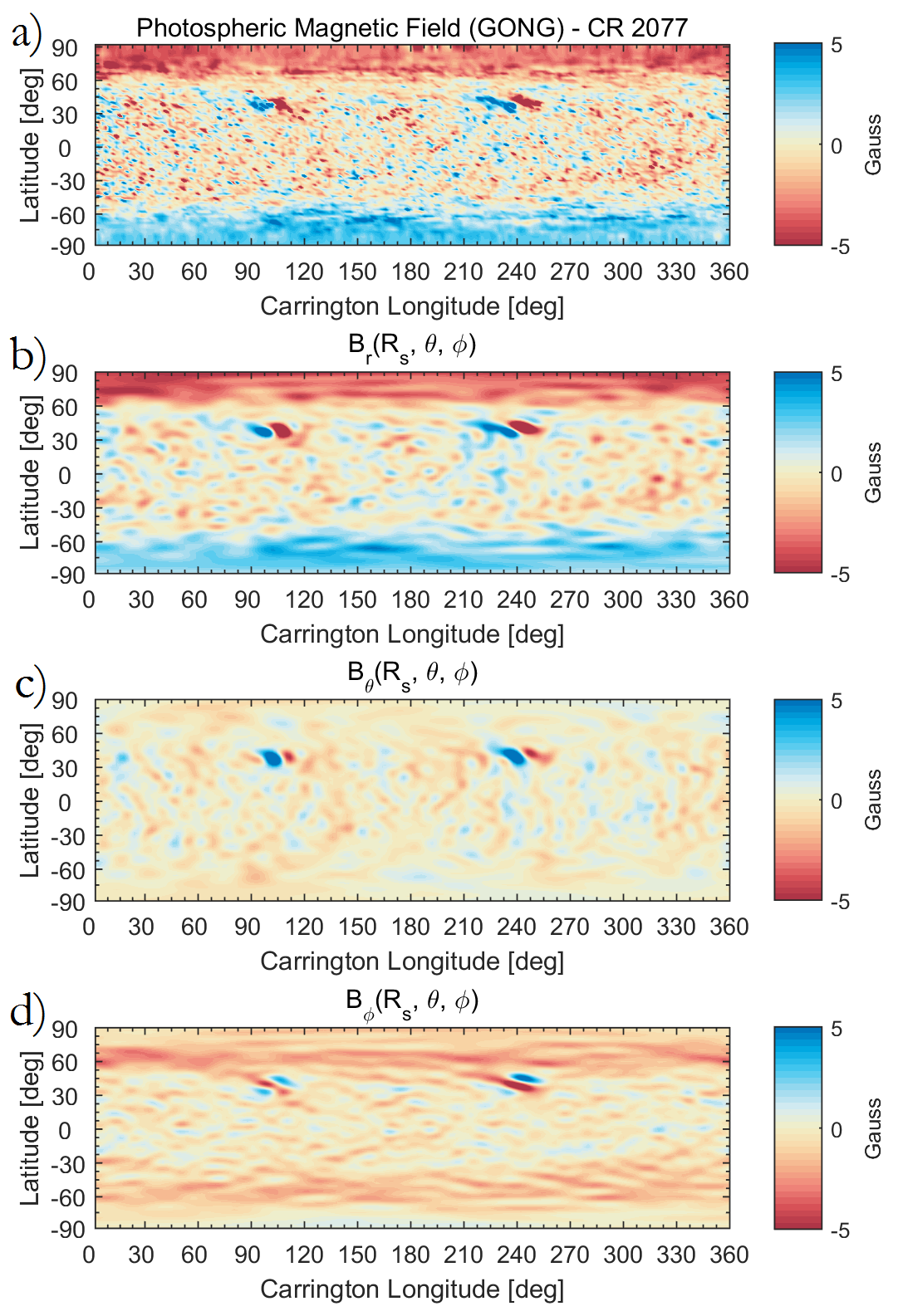}
\end{center}
\caption{Comparison of the observed line-of-sight component of the photospheric field and the PFSS model solution as a function of the heliographic latitude and Carrington longitude for CR2077 (i.e., 2008 November~20--December~17). {The magnetic field strength is saturated at $\pm 5$ G.} (a) Remeshed GONG synoptic magnetogram with $N_\theta = 181$; (b)--(d) Magnetic field solutions $B_\text{r}$, $B_\theta$, and $B_\phi$ computed from the PFSS model using the spherical harmonics expansion up to the order of 80 spherical harmonics. \label{fig:figure3}}
\end{figure}

In this paper, we present an implementation of a numerical framework for operating, validating, and optimizing models of the evolving ambient solar wind. To study a large number of initial conditions, we rely on the coupling between well-established coronal models~\cite{wang90,riley01,arge03} and the Heliospheric Upwind eXtrapolation (HUX) model~\cite{riley11b} that bridges the gap of ballistic mapping and MHD modeling. While this paper outlines the breakdown structure and the mathematical foundation of the operational framework, a subsequent paper will be devoted to the validation and optimization of the numerical models by examining the relationship between the coronal magnetic fields and properties of the ambient solar wind. This paper is organized into three sections. In Section~\ref{sec:2}, we discuss the numerical framework for forecasting the ambient solar wind, including the remeshing of magnetograms (Section~\ref{sec:21}), the magnetic model of the solar corona (Section~\ref{sec:22}), the features of the coronal field solution (Section~\ref{sec:23}), the empirical relationships for specifying the solar wind conditions near the Sun (Section~\ref{sec:24}), the mapping of the solar wind solutions to Earth (Section~\ref{sec:25}), the application of a sensitivity analysis (Section~\ref{sec:26}), and the quantification of the forecasting performance (Section~\ref{sec:27}); and in Section~\ref{sec:3} we conclude with a summary of the results and outline future applications of the framework. 

\section{Description of the Numerical Framework} \label{sec:2}

We present an implementation of a numerical framework for operating, validating, and optimizing models of the ambient solar wind. Figure~\ref{fig:figure2} illustrates the breakdown structure of the current framework implementation. The coronal domain spans the range from $1 \,R_{0}$ to $5 \,R_{0}$, where the outer boundary condition computed from the coronal part is used as an inner boundary condition for the heliospheric domain. The model in the heliospheric domain where the solar wind flow is supersonic then uses the boundary condition as an input for propagating the solar wind solutions near the Sun to 1~au. The subsequent sections are concerned with explaining the components of the framework in detail.

\subsection{Input to the Models} \label{sec:21}
We use full-disk photospheric magnetograms from the Global Oscillation Network Group (GONG) from the National Solar Observatory (NSO) as an inner boundary condition for the coronal model. The global maps of the solar magnetic field measured in Gauss (G) are given on the $\left(\sin \theta, \phi\right)$ grid with $180 \times 360$ grid points, where $\theta \in [0,\pi]$ and $\phi \in [0,2\pi]$ are the latitude and longitude coordinates, respectively. They are available as near-real-time magnetic maps or full CR maps at the GONG online platform\footnote{\url{http://gong.nso.edu}}. Throughout this paper, we illustrate our model solutions on the example of CR2077 (2008 November~20--December~17), that is, during solar minimum when the global magnetic field is dominated by its dipolar component and pronounced polar coronal holes are observable.

We use the full-disk magnetic maps as an inner boundary condition for the coronal model. The coronal model is based on a spherical harmonic decomposition of the input magnetogram. When the raw magnetogram on the $\left(\sin \theta, \phi\right)$ grid is used as an input and the order of spherical harmonics expansion is large compared to the resolution of the magnetic map, inaccurate results, especially at the crucial polar regions, are expected. \citet{toth11} concludes that the information from magnetograms is used more efficiently when the input magnetograms (i) are remeshed to a grid that is uniform in the latitude $\theta$, (ii) contain both poles at $\theta=0$, and $\theta=\pi$, and (iii) have an odd number of grid points. 

We remesh the magnetograms according to the linear interpolation procedure discussed in~\citet{toth11}. To do so, we begin with adding two additional grid cells at the north and south poles of the input magnetic map $M_{i,j}$ with $N_\theta=180$ and $N_\phi=360$ grid points in the latitude and longitude, respectively. The values for the extra grid cells at the poles $M_0$ and $M_{N_\theta +1}$ are given by

\begin{equation}
M_0 = \frac{1}{N_\phi} \sum_{j=1}^{N_\phi} M_{1,j},
\label{eq:}
\end{equation}
and

\begin{equation}
M_{N_\theta + 1} = \frac{1}{N_\phi} \sum_{j=1}^{N_\phi} M_{N_\theta,j}.
\label{eq:}
\end{equation}
The latitude of the uniform $\theta$ grid is

\begin{equation}
\theta_{i'}' = \pi \frac{i' - 1}{N'_\theta - 1}, 
\label{eq:}
\end{equation}
where $N'_\theta$ is the number of grid points at the uniform $\theta$ grid, and the index $i' = 1 \dots N'_\theta$. Finally, we interpolate the grid points from the raw magnetogram mesh to the uniform $\theta$ mesh using a linear interpolation relation (no magnetic flux conservation) of the form

\begin{equation}
M'_{i',j} = \alpha M_{i,j} + (1- \alpha) M_{i+1, j},
\label{eq:}
\end{equation}
where the index $i$ is selected so that 

\begin{equation}
\theta \le \theta'_{i'} \le \theta_{i+1},
\label{eq:}
\end{equation}
and 

\begin{equation}
\alpha = \frac{\theta_{i+1} - \theta'_{i'}}{\theta_{i+1} - \theta_i}.
\label{eq:}
\end{equation}
Notice that the maximum degree for the expansion of spherical harmonics is now limited only by the anti-alias limit~\cite[see,][]{toth11} given by

\begin{equation}
\min\left(\frac{2 N_\theta'}{3}, \frac{N_\phi}{3}\right).
\label{eq:}
\end{equation}
By utilizing remeshed magnetograms as an inner boundary condition for the coronal model, we can, therefore, expect accurate solutions up to the order of 120 spherical harmonics~\cite{toth11}. Figure~\ref{fig:figure3}(a) depicts the computed remeshed magnetogram as a function of the heliographic latitude and Carrington longitude for CR2077. 

\subsection{Magnetic Model of the Solar Corona} \label{sec:22}
The magnetic model of the solar corona couples the PFSS model and the SCS model to reconstruct the coronal magnetic field. The PFSS model is based on the assumptions that the region above the photosphere is current free, and that the magnetic field at an imaginary reference sphere, called the source surface with radius $R_1$ is radial only~\cite{altschuler69,schatten69}. As detailed in Appendix~\ref{app:pfss}, Eq.(\ref{eq:A4}) can be solved to derive the magnetic field components at any point in the coronal domain ($R_0 \le r \le R_1$). Figure~\ref{fig:figure3}~(b)--(d) present the magnetic field components $B_\text{r}, B_\theta, and B_\phi$ derived from the PFSS model at the solar surface as a function of the latitude and Carrington longitude for CR2077. {We note that we used a color scheme for all illustrations that is color-blind friendly~\cite[see,][]{light04}.} From the comparison in Figure~\ref{fig:figure3}, it is apparent that the PFSS solution using the spherical harmonics expansion up to $n=80$, is in reasonable agreement with the observed line-of-sight component of the photospheric magnetic field.

\begin{figure*}
\begin{center}
\includegraphics[width=0.99\columnwidth]{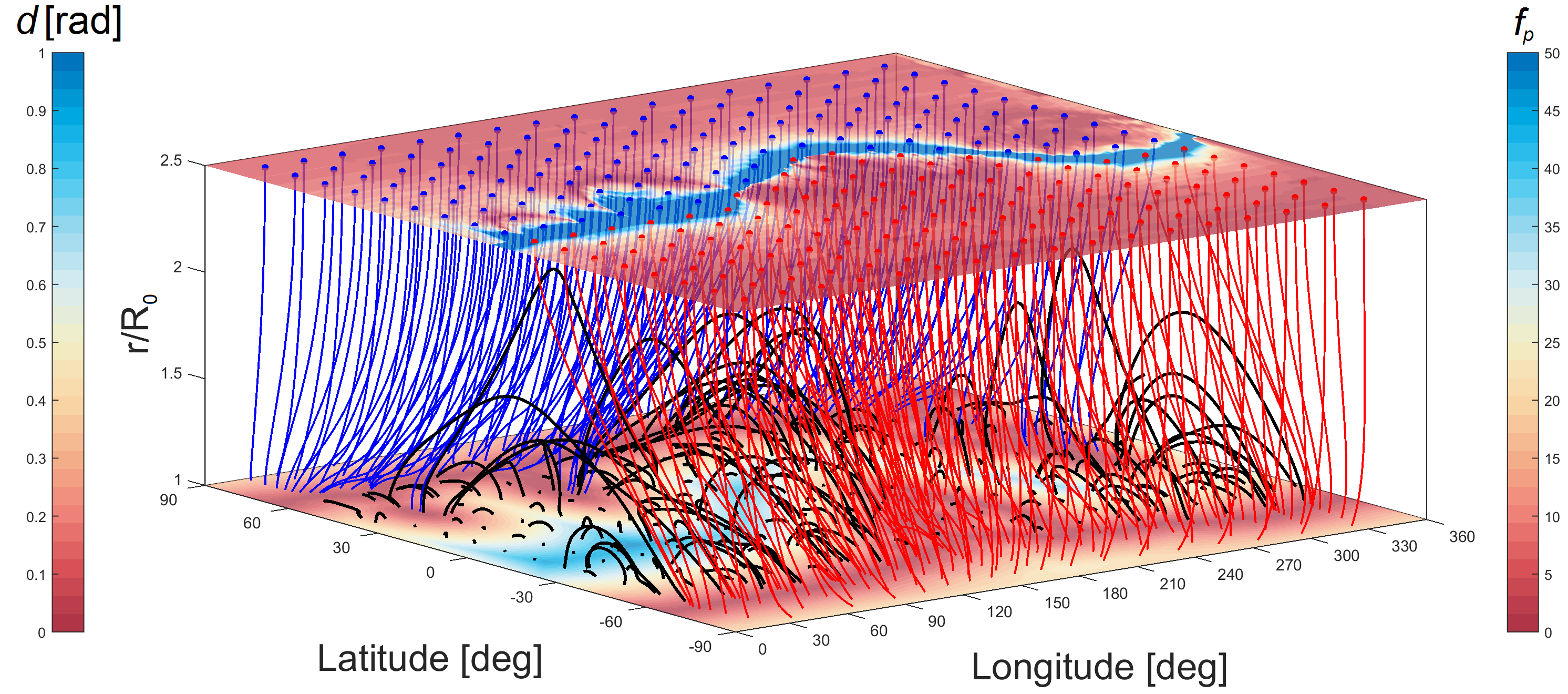}
\caption{Illustration of the topology of the magnetic field between the photosphere and the source surface at $2.5 \,R_0$ computed from the described magnetic model of the solar corona for CR2077. Red and blue colors indicate positive and negative magnetic field lines, respectively. While the lower plane shows the great-circle angular distance from the nearest coronal hole boundary at the photosphere $d$, the upper plane shows the areal expansion factor $f_\text{p}$ indicating how much the corresponding flux tube expands between the photosphere and the source surface at $2.5 \,R_0$. 
\label{fig:figure4}}
\end{center}
\end{figure*}

To extend the magnetic field solution from the outer boundary of the PFSS model at $2.5 \,R_0$ to a distance of $R_2 = 5 \,R_{0}$, we employ the SCS model~\cite{schatten71}. The SCS model uses the PFSS solutions at the source surface as an inner boundary condition. The SCS model is similar to the PFSS model but involves solving an additional Laplace equation with different boundary conditions. In the first step, the magnetic field on the source surface is oriented to point away from the Sun. In other words, the signs of the magnetic field components $B_\text{r}$, $B_\theta$, and $B_\phi$ are reversed if $B_\text{r} < 0$ at the source surface. While the magnetic field solution is defined to vanish at infinity, the non-negative magnetic field values as model input ensure that a thin current sheet is retained in the field solution. The underlying idea of the SCS model is to extend the magnetic field solution in a nearly radial way. To retain the resolution, we match the grid of the SCS model with the PFSS model with equal steps of $2 \degree$ in latitude and longitude. As discussed in Appendix~\ref{app:scs}, we use estimates of the coefficients $g_n^m$ and $h_n^m$ from the least mean square fit to compute the magnetic field above the source surface. Finally, the polarity needs to be corrected to match the polarities in the region $r \le R_1$. 

The imposed boundary conditions of the PFSS model and the SCS model are not compatible, causing discontinuities in the tangential component of the magnetic field across the source surface. When coupling the PFSS model and SCS model solutions, discontinuities in the form of kinks in the magnetic field topology are observable. In order to account for this effect,~\cite{mcgregor08} proposed a more flexible coupling between the two models. The authors concluded that their approach leads to more accurate forecast results. Following the procedure of~\cite{mcgregor08}, we set the radius of the source surface to $2.5 \,R_0$ but use the PFSS solution at $2.3 \,R_0$ as an inner boundary condition for the SCS model. By doing so, we couple the PFSS and SCS model solutions and reconstruct the global topology of the coronal magnetic field. 

\subsection{Computing Features of the Coronal Field Solution} \label{sec:23}

Models for specifying the solar wind conditions near the Sun rely on the areal expansion factor $f_\text{p}$ and the great-circle angular distance from the nearest open-closed boundary $d$, both of which are derived from the coronal magnetic field solution. It is noteworthy that both features are linked to different fundamental theories on the origin of slow solar wind~\cite[see,][]{riley12, riley15}. While the expansion factor assumes that the solar wind flow along open field lines that diverge the most leads to the slow solar wind~\cite{wang90}, the distance to the coronal hole boundary is more related to the boundary layer idea of interchange reconnection for the origin of the slow solar wind~\cite{antiochos11}.

To trace magnetic field lines and reconstruct the magnetic field topology in the coronal model domain $R_0 \le r \le R_2$, we use a fourth-order Runge-Kutta method (RK4) and solve the following set of equations, expressed as

\begin{align}
\begin{split}
\frac{dr}{ds} &= \frac{B_\text{r}}{B} , \\
\frac{d\theta}{ds} &= \frac{1}{r} \frac{B_\theta}{B} , \\
\frac{d\phi}{ds} &= \frac{1}{r \sin \theta} \frac{B_\phi}{B},
\label{eq:}
\end{split}
\end{align}
where $ds$ is a segment along the magnetic field line. 

More specifically, our approach for constructing the large-scale topology of the coronal field is two-fold. First, we start from the base of the model at the solar surface ($r=R_0$) and trace magnetic field lines upwards. When the field line returns to the solar surface, we label the corresponding footpoint at the solar surface as a ``closed field''. In contrast, when the field line reaches the upper boundary of the coronal model ($r=R_2$) we label the footpoint as a ``open field''. In this way, we compute coronal hole regions, i.e., magnetically open regions expected to guide high-speed solar wind streams out into the heliosphere. Using the location of coronal holes at the solar surface, we run a perimeter tracing algorithm to compute coronal hole boundaries. Then, we compute the great-circle angular distance $d$ between footpoints of open field lines and their nearest coronal hole boundary. Second, we start from the outer boundary of the coronal model ($r = R_2$) and trace the field lines down to the solar surface. By doing so, we compute the areal expansion factor $f_\text{p}$ as

\begin{align}
f_\text{p} = \left( \frac{R_0}{R_1} \right)^2 \left|\frac{B_\text{r} (R_0, \theta_0, \phi_0)}{B_\text{r} (R_1, \theta_1, \phi_1)}\right|,
\end{align}
where $B_\text{r}$ is the radial magnetic field component at a given reference height. The areal expansion factor is the amount by which a flux tube expands from the solar surface to another reference height in the corona. Figure~\ref{fig:figure4} shows the topology of the coronal magnetic field and illustrates the features $d$ and $f_\text{p}$ at the photosphere and the source surface, respectively. 

\begin{figure*}
\begin{center}
\includegraphics[width=0.99\columnwidth]{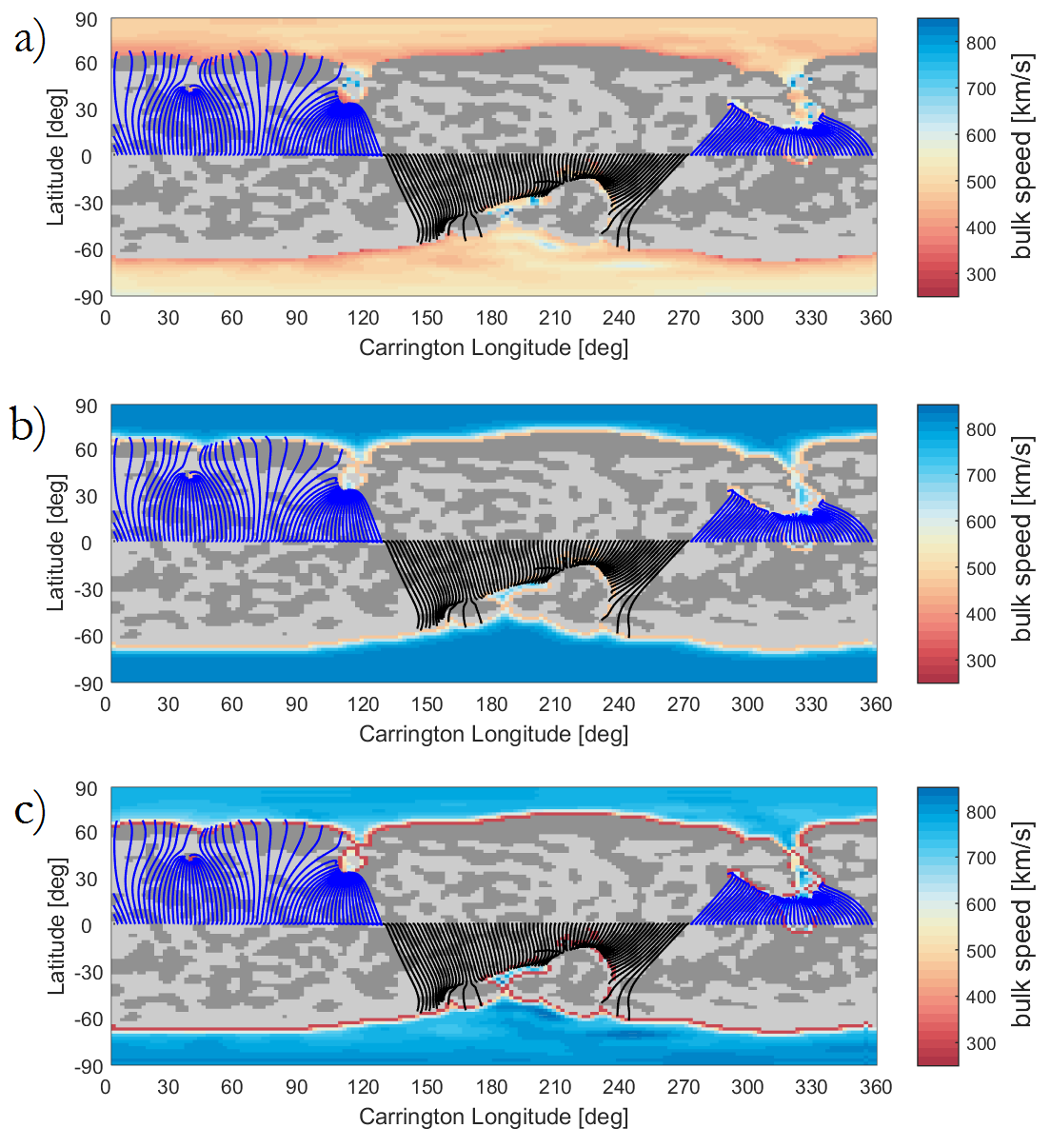}
\caption{Comparison of empirical relationships for specifying the solar wind speed near the Sun on the example of CR2077. The magnetic field lines illustrate the magnetic connectivity between the computed coronal hole regions and the sub-Earth orbit at the outer boundary of the coronal model ($5 \,R_0$). The gray colored pixels indicate closed magnetic field topology with negative (dark gray) and positive (light gray) polarity. The full black and blue lines indicate positive and negative magnetic field lines, respectively. (a) Wang-Sheeley (WS) Model; (b) Distance from the Coronal Hole Boundary (DCHB) Model; (c) Wang-Sheeley-Arge (WSA) Model.\label{fig:figure5}}
\end{center}
\end{figure*}

\subsection{Computing Solar Wind Speed Maps} \label{sec:24}
The structure in the solar wind flow is governed by the dynamic pressure term in the momentum equation ($\propto \rho \, v^2$). This indicates that the near-Sun solar wind solution used as an inner boundary condition for heliospheric models is highly sensitive to errors in the computed solar wind bulk speed~\cite{riley15}. 
Recently,~\citet{riley15} highlighted three empirical relationships for specifying the solar wind speed $(v=v(d,f_\text{p}))$ at a reference sphere of $5 \,R_{0}$ (or $30 \,R_{0}$ for the MAS model). In the literature, these empirical relationships are known as the Wang-Sheeley (WS) model~\cite{wang90}, the Distance from the Coronal Hole Boundary (DCHB) model~\cite{riley01}, and the Wang-Sheeley-Arge (WSA) model~\cite{arge03}. 

The WS model is based on the inverse relationship between the solar wind speed and the magnetic field expansion factor~\cite{wang90}, namely

\begin{align}
v_{\text{ws}} (f_\text{p}) = v_{0} + \frac{(v_{1} - v_{0})}{{f_\text{p}}^\alpha}.
\end{align}
Previous research has established that low magnetic field expansion between the photosphere and some reference height in the corona is correlated with a fast solar wind speed~\cite[e.g.,][]{levine77}. For the coefficients in Eq.(10), we use $v_0=250$, $v_1=660$, and $\alpha=2/5$~\cite[see,][]{arge00}.
 
The DCHB model relates the speed at the photosphere with the distance of an open field footpoint from the coronal hole boundary and maps the calculated solar wind speed solution out along the magnetic field lines to a given reference sphere~\cite{riley01}. The DCHB relation is defined as

\begin{align}
v_{\text{dchb}} (d) = v_{0} + \frac{1}{2} \left(v_{1} - v_{0}\right) \left[1 + \tanh \left( \frac{d - \epsilon}{w} \right)\right],
\end{align}
where $\epsilon$ is a measure for the thickness of the slow flow band, and $w$ denotes the width over which the solar wind reaches coronal hole values. For an open field footpoint at the coronal hole boundary, the solar wind speed is equal to $v_0$. For a footpoint located deep inside a coronal hole, the solar wind speed is equal to $v_1$. Hence, the farther away the footpoint is from a coronal hole boundary, the faster the expected solar wind speed. In this study, we use $v_0=350$, $v_1=750$, $\epsilon=0.1$, and $w=0.05$~\cite[see,][]{riley01}.

Finally, the WSA model is a hybrid of the WS model and the DCHB model, combining aspects of the expansion factor computed from the topology of the magnetic field and the distance to the coronal hole boundary~\cite{arge03}. The WSA model for specifying solar wind speed is given by 

\begin{align}
v_{\text{wsa}} (f_\text{p}, d) = v_{0} + \frac{\left(v_{1} - v_{0}\right)}{\left(1 + f_\text{p}\right)^\alpha} \left\{ \beta - \gamma \ \exp \left[ {-(d/w)}^{\delta}\right] \right\}^3,
\end{align}
where $\alpha$, $\beta$, $\gamma$, and $\delta$ are additional model parameters. For the coefficients in Eq.(12) we use $v_0=285$, $v_1=910$, $\alpha=2/9$, $\beta=1$, $\gamma=0.8$, $w=2$, and $\delta =3$. Figure~\ref{fig:figure5} shows a comparison between the different empirical relationships for specifying a solar wind speed of $v(f_\text{p},d)$ near the Sun.

\subsection{Mapping of the Solar Wind to Earth} \label{sec:25}

\begin{figure*}
\begin{center}
\includegraphics[width=0.99\columnwidth]{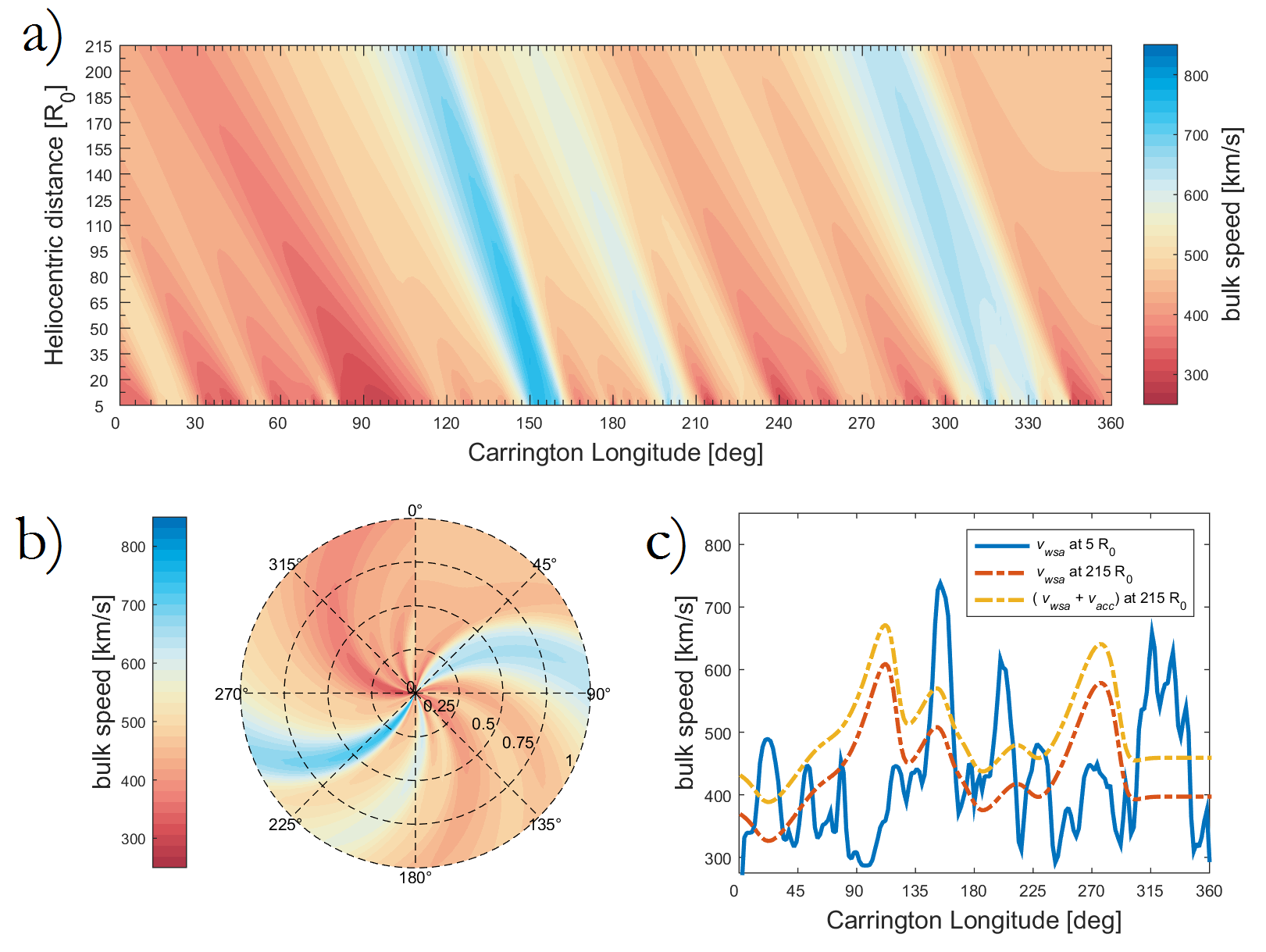}
\caption{The Heliospheric Upwind eXtrapolation (HUX) model for mapping the WSA solar wind speed solution to 1~au (or $215 \,R_0$). (a)--(b) Solar wind bulk speed as a function of the heliocentric distance. (c) The input solar wind speed at $5 \,R_0$ (blue), and the resulting solar wind speed at $215 \,R_0$ with (yellow) and without (red) the residual speed contribution $v_{\text{acc}}$.
\label{fig:figure7}}
\end{center}
\end{figure*}

Numerous models dealing with the mapping of solar wind solutions near the Sun to Earth have been developed. The spectrum extends from a simple ballistic approximation where each parcel of plasma is assumed to travel with a constant speed through the heliosphere to global heliospheric MHD models that aim to cover all relevant dynamical processes~\cite[e.g.,][]{riley11c,odstrcil03}. In an attempt to bridge the gap,~\citet{riley11b} developed the Heliospheric Upwind eXtrapolation (HUX) model. \citet{riley11b} have simulated the kinematic of the solar wind flow numerically by simplifying the MHD equations as much as possible. By neglecting the pressure gradient and the gravity terms in the fluid momentum equation~\cite[e.g.,][]{pizzo78} the solar wind solution on a discretized grid may be written as
\begin{eqnarray}
v_{r+1,\phi} = v_{r,\phi} + \frac{\Delta r \ \Omega}{v_{r,\phi}} \left( \frac{v_{r,\phi+1} - v_{r,\phi} }{\Delta \phi} \right),
\end{eqnarray}
where the subscripts $r$ and $\phi$ refer to radial and longitudinal grid cells with cell spacings $\Delta r$ and $\Delta \phi$, and $\Omega$ denotes the equatorial angular rotation rate of the Sun (neglecting differential rotation)~\cite[see,][]{riley11b}. In order to match the spatial resolution of the coronal model, we use $\Delta \phi = 2\degree$ and $\Delta r = 1 \,R_0$, respectively. As the step size in $\Delta r$ is increased by two orders of magnitude, the solutions are not significantly different, indicating that the Courant-Friedrichs-Lewy condition for numerical convergence is well met.

Furthermore,~\citet{riley11b} concluded that it is convenient to account for the residual wind acceleration beyond the coronal model. According to previous model results~\cite[e.g.,][]{riley01, riley11b}, the residual acceleration $v_{\text{acc}}$ expected beyond the coronal model is 
\begin{eqnarray}
v_{\text{acc}}(r) = \alpha \ v_{r_0} \left[ 1 - \exp \left(\frac{-r}{r_h}\right) \right],
\end{eqnarray}
where $v_{r_0}$ is the initial speed at the outer boundary of the coronal model, $r$ is the heliocentric distance, $\alpha$ is a factor by which $v_{r_0}$ is enhanced, and $r_h$ is the scale length for the acceleration (not crucial when mapping the solar wind to 1~au). For the coefficients in Eq.(14) we use $\alpha=0.15$ and $v_{r_0}=50 \,R_0$. 

Figure~\ref{fig:figure7} illustrates the HUX model for mapping the solar wind solution near the Sun to Earth. While the top panel shows the propagation of the solar wind solutions in steps of $\Delta r$ from $5 \,R_0$ to $215 \,R_0$ (or 1~au), the bottom panel shows the propagation in spherical coordinates (left) and the calculated speed time series at Earth with and without the residual speed contribution (right). It is also noteworthy that computing the solar wind solution for other heliospheric distances is straightforward. The benefit of the HUX model is that it can match the dynamical evolution explored by global heliospheric MHD models that demand only low computational requirements~\cite{riley11b}. This is particularly useful to study a large number of initial conditions in the context of ensemble modeling. 

\begin{figure*}
\begin{center}
\includegraphics[width=0.99\columnwidth]{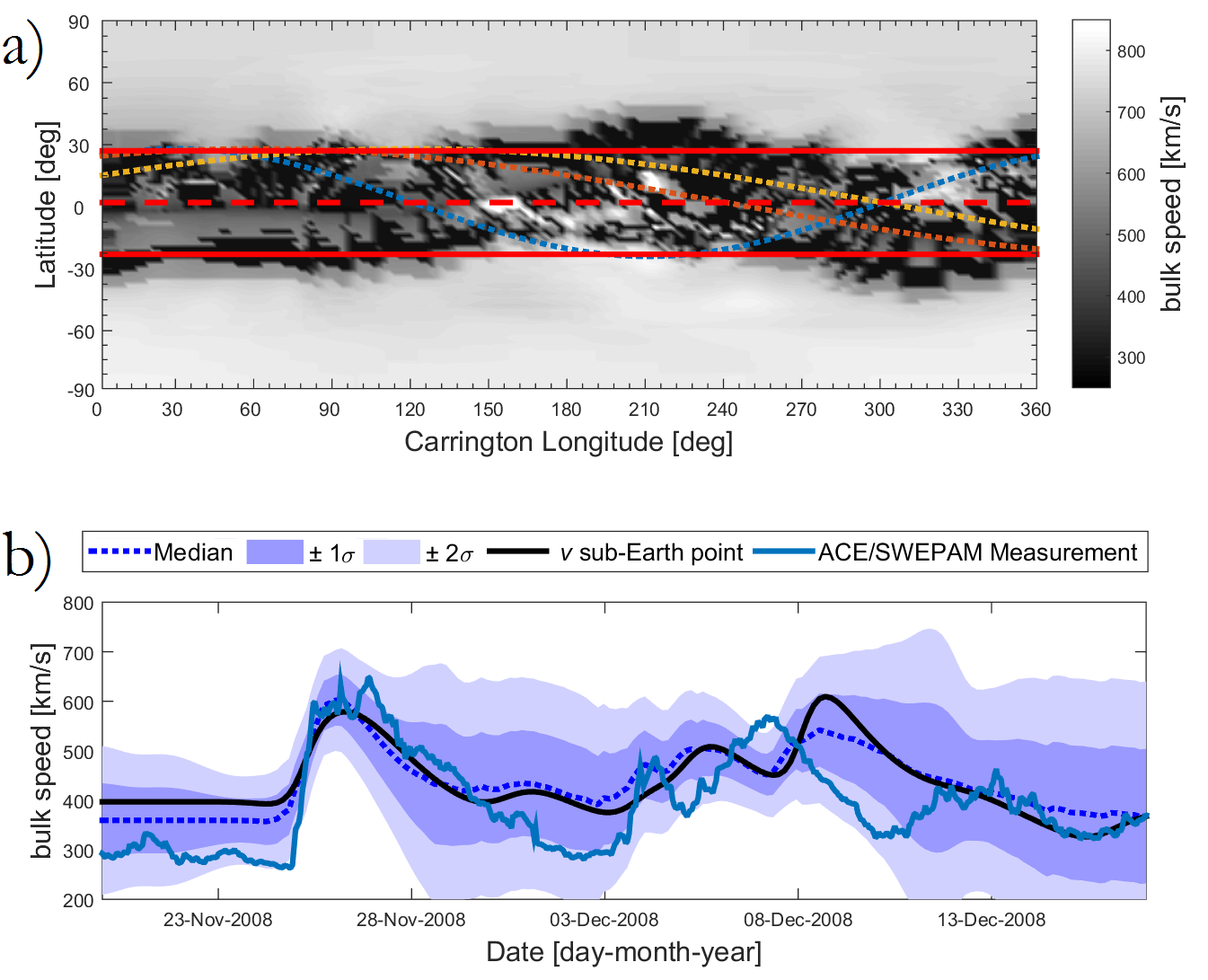}
\caption{Illustration of the ensemble of initial conditions and the process of mapping the solar wind solutions to Earth by the HUX model. (a) Solar wind solution for CR2077 from the WSA model at $5 \,R_0$ overlaid with 3 out of 576 individual trajectories together with the median value (dashed red line) and the $\pm 2\,\sigma$ quantiles (red line); (b) A comparison of observed and WSA model solar wind speed time series at $215 \,R_0$. {The ensemble median of solar wind solutions} at Earth is indicated by the dashed blue line and the $\pm 2\,\sigma$ quantiles are shown in blue color.
\label{fig:figure8}}
\end{center}
\end{figure*}

\begin{table*}[t]
\caption{The skill of model predictions of solar wind speed for CR2077 in terms of Arithmetic Mean (AM), Standard Deviation (SD), Mean Error (ME), Mean Absolute Error (MAE), and Root-Mean-Square Error (RMSE). A 4-day and 27-day persistence model of near-Earth solar wind conditions provides a baseline against which solar wind models can be compared.}
\begin{center}
\begin{tabular}{lccccccc}
Model         & \multicolumn{1}{l}{AM {[}km/s{]}} & \multicolumn{1}{l}{SD {[}km/s{]}} & \multicolumn{1}{l}{ME {[}km/s{]}} & \multicolumn{1}{l}{MAE {[}km/s{]}} & \multicolumn{1}{l}{RMSE {[}km/s{]}} \\ \hline
WS            & 428.79                              & 62.63                             & -27.90                            & 74.09                              & 85.27                               \\
DCHB          & 379.47                              & 30.93                             & 21.42                             & 83.83                              & 103.43                              \\
WSA           & 437.14                              & 69.14                             & -36.25                            & 68.54                              & 82.62                               \\
{Ensemble Median (WS)}   & 435.55                              & 51.84                             & -34.66                            & 71.52                              & 83.36                               \\
{Ensemble Median (DCHB)} & 367.47                              & 4.62                              & 33.43                             & 78.27                              & 100.04                              \\
{Ensemble Median (WSA)}  & 437.48                              & 63.64                             & -36.56                            & 62.24                              & 74.86                               \\
Persistence (4-days)  & 394.76                            & 99.23                             & 6.13                           & 130.48                              & 161.99                               \\
Persistence (27-days)  & 409.94                              & 108.91                            & -9.05                            & 66.54                              & 78.86                              \\
Observation   & 400.89                              & 96.12                             & -                                 & -                                  & -                                  
\end{tabular}
\end{center}
  \label{tab:1}
\end{table*}

\subsection{Creating Ensembles of Initial Conditions for Solar Wind Forecasting} \label{sec:26}

The knowledge on the sensitivity of the model performance to initial conditions and model parameter settings is crucial for models of the ambient solar wind. In recent years, the concept of ensemble forecasting has been successfully applied to a number of applications in the space weather forecasting regime. As an example, the sensitivity of PFSS and MHD coronal models on magnetic maps from different observatories has been studied by~\citet{riley13a,riley13b}. More details on recent activities can be found in the reviews by~\citet{knipp16} and~\citet{murray18}. Ensemble forecasting is a technique based on the use of a sample of possible future states to forecast a future state. Intuitively, one would expect that the forecast of the ambient solar wind conditions is very uncertain when the ensemble members (e.g., different solar wind models) are very different from one another. The strength of ensemble forecasting is its capability to deduce confidence bounds by quantifying the uncertainty in the ensemble of possible future states. 

\citet{owens17}, for instance, has pointed out that the forecast uncertainty in solar wind models can be studied by adding uncertainty in the sub-Earth orbit at which the coronal solutions near the Sun are sampled. Using the approach of~\citet{owens17}, we perform a sensitivity analysis that considers ensemble members at perturbed latitudes $\theta_p$ around the sub-Earth orbit given by
\begin{eqnarray}
\theta_p (\phi) = \theta_E + \theta_{1} \, \sin \left( n \,\phi + \phi_0 \right),
\end{eqnarray}
where $\phi$ is the corresponding Carrington longitude; $\theta_E$ is the sub-Earth latitude; $\theta_{1}$ is the amplitude of the deviation; $n$ is the wave number, indicating how many oscillations the perturbed trajectory completes; and $\phi_0$ is the phase offset, indicating how two perturbed trajectories can be out of synchronicity with each other. For the coefficients in Eq.(14), we select $\theta_{1} \in [0 \degree, 15 \degree]$ in $1 \degree$ steps, $n \in [0, 1]$ in $0.5$ steps, and $\theta_{0} \in [0 \degree, 330 \degree]$ in $30 \degree$ steps, which ensures that all latitudes are well represented at each longitude~\cite[see,][]{owens17}. 

Figure~\ref{fig:figure8}(a) shows three individual trajectories together with the median value and the $\pm 2 \sigma$ quantiles of the ensemble members. In this way, we obtain information from a sampling of 576 initial speed solutions, each of which are mapped to Earth using the HUX model as outlined in Section~\ref{sec:25}. Figure~\ref{fig:figure8}(b) compares the resulting {ensemble median} at $215 \,R_{0}$ and the observed solar wind speed in the near-Earth environment. {In this study, the ensemble median is the preferred average measure as the ensembles of solar wind solutions around the sub-Earth latitude are often highly skewed, such that the ensemble mean can yield very biased measures of the ensemble average. Further, we note that ensemble averaging of possible future states is not necessarily expected to provide a systematic improvement of deterministic forecasts~\cite[see,][]{jolliffe03}. As discussed in \citet{owens17} and \citet{henley17}, it is the assessment of forecast uncertainty that is the key advantage of the ensemble approach rather than the ensemble average providing an improved deterministic forecast.}

\subsection{Assessing the Forecast Performance} \label{sec:27}

The information from the implemented models is used most efficiently when the uncertainty of their results is constantly validated. To enable that, we measure the performance of the framework by the Operational Solar Wind Evaluation Algorithm~\cite[OSEA;][]{reiss16}. OSEA\footnote{\url{https://bitbucket.org/reissmar/solar-wind-forecast-verification}} runs various validation procedures to compare the forecasts and the observations to which they pertain. Traditionally, the relationship between forecast and observation can be studied in terms of continuous variables and binary variables. While the former can take on any real values, the latter is restricted to two possible values such as event/non-event. In the context of solar wind forecasting, the solar wind speed time series can be interpreted in terms of both aspects. The forecasting performance can either be evaluated in terms of an average error or in terms of the capability of forecasting events of an enhanced solar wind speed~\cite{owens05,macneice09a,macneice09b}. OSEA is capable of quantifying both aspects, i.e., a continuous variable evaluation that uses simple point-to-point comparison metrics and an event-based validation analysis that assesses uncertainty of the arrival time of high-speed solar wind streams at Earth.

As an example, Table~\ref{tab:1} lists the results obtained from the continuous variable validation of different solar wind models for CR2077 in terms of Arithmetic Mean (AM), Standard Deviation (SD), Mean Error (ME), Mean Absolute Error (MAE), and Root-Mean-Square Error (RMSE). A 4 day and 27 day persistence model of near-Earth solar wind conditions provides a baseline against which the performance can be compared. We find that the RMSE for the WS model, DCHB model, and WSA model is 85.27~km/s, 103.43~km/s, and 82.62~km/s, respectively. The 27 day persistence model has the same statistics as the measurements and greatly benefits from the quasi-steady and recurrent nature of the evolving ambient solar wind in the solar minimum phase and thus is expected to score very high in all measures (e.g., $\text{RMSE} = 78.86$~km/s). We conclude that the WSA model gives the best forecast results in terms of a continuous variable validation and that the process of ensemble forecasting slightly improves the performance of all solar wind models in this study. It is important to note that the present analysis is indented to illustrate the application of the implemented framework components, which means that our results apply only to CR2077 and that our conclusions are not reliable as a general guide.

\section{Discussion} \label{sec:3}
We present a numerical framework for forecasting the evolving ambient solar wind that uses magnetic maps as an input for the PFSS and SCS model to reconstruct the global topology of the coronal magnetic field, specifies the solar wind speed using different established empirical relationships (WS, DCHB, and WSA models) based on the areal expansion factor and the distance to the nearest coronal hole boundary, maps the near-Sun solar wind solution outward to Earth by the HUX model, creates an ensemble of initial conditions by adding uncertainty in the latitude about the sub-Earth point, and uses an automated forecast validation module to quantitatively assess the forecasting skill. The framework relies on established models of the ambient solar wind and is conceptually very similar to already existing numerical frameworks. We carefully compared our model solutions to existing frameworks~\cite[e.g.,][]{nikolic14,arge03}, and found that our modular framework implementation in the C\texttt{++} and Matlab programming language {using tools from the Armadillo library~\cite{sanderson16}} is both robust and fast. 

The coronal part of the framework relies on empirical relationships between the magnetic field topology and the near-Sun solar wind conditions~\cite{wang90,riley01,arge03}. Ideally, one would prefer the application of a physics-based MHD model for the corona to capture the complex dynamics at solar wind stream interaction regions, which are not included in the described model approach. Nevertheless, recent studies have shown that the forecast skill of empirical and full physics-based coupled corona-heliosphere models in terms of established metrics is very similar. As an example, \cite{owens08} studied the performance of different forecast models (WSA, WSA-Enlil, and MAS-Enlil) over an 8~yr period and concluded that the coupled empirical approach currently gives the best forecast results in terms of the mean square error. Considering the trade-off between accuracy and computational requirements of a full MHD code, it thus seems reasonable to follow the described methodology in the context of solar wind forecasting. We conclude that the efficient implementation of the framework using the heliospheric HUX model is well suited for studying the long-term relationship between coronal magnetic fields and the properties of the ambient solar wind. 

In the future, we shall work on several topics to try to improve the forecasting performances. While this study presents the implementation of the numerical framework, a subsequent paper will be devoted to the validation and optimization of the present solar wind models. By studying the coupling between magnetic models of the corona and those of the inner heliosphere, we aim toward optimizing the empirical relationships for specifying solar wind properties to advance further their predictive capabilities. In context, a reliable forecast of the ambient solar wind might be beneficial for forecasting the arrival of CMEs. Here, we plan to combine the ambient solar wind framework with the ELlipse Evolution model based on Heliospheric Imager observations~\cite[ELEvoHI;][]{rollett16,amerstorfer18}. The kinematics of the elliptical-shaped CME front in the ELEvoHI model is governed by the ambient solar wind flow. In the current version of ELEvoHI, the solar wind speed is assumed to be constant during the propagation of the CME in the inner heliosphere. Therefore, we would also like to include information on the complex dynamics of the evolving ambient solar wind to simulate the dynamic deformation of the CME front during the propagation phase. To make the model runs accessible to the space weather community, we also plan to install a later version of the solar wind framework at NASA's CCMC online platform. {The release of this online resource is scheduled for 2019 July.}

\section*{Acknowledgments}
The work utilizes data obtained by the Global Oscillation Network Group (GONG) Program, managed by the National Solar Observatory, which is operated by AURA, Inc.~under a cooperative agreement with the National Science Foundation. The data were acquired by instruments operated by the Big Bear Solar Observatory, High Altitude Observatory, Learmonth Solar Observatory, Udaipur Solar Observatory, Instituto de Astrof\'{i}sica de Canarias, and Cerro Tololo Interamerican Observatory. L.N.~performed  this  work as part of Natural Resources Canada's Public Safety Geoscience program. M.A.R., C.M.~and T.A.~acknowledge the Austrian Science Fund (FWF): J4160-N27, P26174-N27 and P31265-N27.

\appendix
\section{PFSS Model} \label{app:pfss}

The PFSS model is based on the assumption that the magnetic field $\mathbf{B}$ above the photosphere is current free ($\nabla \times \mathbf{B}= 0$). With this approximation, the magnetic field can be expressed as the gradient of a scalar potential $\Psi$,

\begin{equation}
\mathbf{B} = - \nabla \Psi.
\label{eq:}
\end{equation}
Using $\nabla \cdot \mathbf{B} = 0$, which expresses the fact that there are no magnetic monopoles, we can write the Laplace equation for the potential $\Psi$ as

\begin{equation}
\nabla^2 \Psi = 0. 
\label{eq:}
\end{equation}
The solution of the Laplace equation in spherical coordinates with $\theta \in [0,\pi]$ and $\phi \in [0,2 \pi]$ in the region $R_0 \leq r \leq R_1$ can be expressed as a infinite series of spherical harmonics expressed as

\begin{equation}
\begin{split}
\Psi (r, \theta, \phi) &= \left[ R_0 \left( \frac{R_0}{r}\right)^{n+1} - R_1 \left( \frac{R_0}{R_1}\right)^{n+2} \left( \frac{r}{R_1}\right)^n \right] \\ &\times \sum_{n = 1}^{\infty}{\sum_{m=0}^n{ \left(g_n^m \cos m \phi + h_n^m \sin m \phi \right) P_n^m (\cos \theta)}},
\label{eq:}
\end{split}
\end{equation}
where $R_1 = 2.5 \ R_{0}$ is the source surface radius, $P_n^m (\cos \theta)$ are the associated Legendre polynomials, and $g_n^m$ and $h_n^m$ are coefficients that are computed from the input magnetograms. Using Eq.(A1), the solution for the magnetic field components is given by

\begin{equation}
\mathbf{B} = (B_\text{r}, B_\theta, B_\phi) = \left( -\frac{\partial \psi}{\partial r}, -\frac{1}{r}\frac{\partial \psi}{\partial \phi}, -\frac{1}{r \sin \theta}\frac{\partial \psi}{\partial \phi}\right).
\label{eq:A4}
\end{equation}
The magnetic field components are computed as follows:

\begin{equation}
\begin{split}
B_\text{r} &= \left[ (n+1) \left( \frac{R_0}{r}\right)^{n+2} + \left( \frac{R_0}{R_1}\right)^{n+2} \left( \frac{r}{R_1}\right)^{n-1} \right] \\ &\times \sum_{n = 1}^{\infty}{\sum_{m=0}^n{ \left(g_n^m \cos m \phi + h_n^m \sin m \phi \right) P_n^m (\cos \theta )  }},
\label{eq:}
\end{split}
\end{equation}
\begin{equation}
\begin{split}
B_{\theta} &= - \left[ \left( \frac{R_0}{r}\right)^{n+2} - \left( \frac{R_0}{R_1}\right)^{n+2} \left( \frac{r}{R_1}\right)^{n-1} \right] \\ &\times \sum_{n = 1}^{\infty}{\sum_{m=0}^n{ \left(g_n^m \cos m \phi  + h_n^m \sin m \phi \right) P_n^m (\cos \theta )}},
\label{eq:}
\end{split}
\end{equation}
\begin{equation}
\begin{split}
B_{\phi} &= \left[ \left( \frac{R_0}{r}\right)^{n+2} - \left( \frac{R_0}{R_1}\right)^{n+2} \left( \frac{r}{R_1}\right)^{n-1} \right] \\ &\times \sum_{n = 1}^{\infty}{\sum_{m=0}^n{ \frac{m}{\sin \theta }\left(g_n^m \sin m \phi  - h_n^m \cos m \phi \right) P_n^m (\cos \theta )}}.
\label{eq:}
\end{split}
\end{equation}
To compute the coefficients $g_n^m$ and $h_n^m$, we multiply Eq.(A3) for $r=R_0$ with $P_{n'}^{m'} \cos m'\phi$ and $P_{n'}^{m'} \sin m'\phi$, respectively. Using the orthogonality of the Schmidt normalized Legendre polynomials and integrating over the spherical surface, we find

\begin{align}
\frac{1}{4\pi} \int_0^{\pi} \int_0^{2\pi} P_n^m(\theta) \begin{Bmatrix} \cos m\phi \\ \sin m\phi \end{Bmatrix} P_{n'}^{m'}(\theta) \begin{Bmatrix} \cos m'\phi \\ \sin m'\phi \end{Bmatrix} \sin \theta \, d\theta \, d\phi = \frac{1}{2n + 1} \delta_n^{n'} \delta_m^{m'}. 
\end{align}
This yields the solution of the coefficients $g_n^m$ and $h_n^m$ in the form
\begin{eqnarray}
\begin{Bmatrix} g_n^m \\ h_n^m \end{Bmatrix} = \frac{2n + 1}{ 4\pi \left(n + 1 + n \left(\frac{R_0}{R_1}\right)^{2n+1}\right)} \int_0^\pi d\theta \, \sin \theta \ P_n^m(\theta) \int_0^{2\pi} d\phi \ B_\text{r}(R_0, \theta, \phi) \begin{Bmatrix} \cos m \phi \\ \sin m \phi \end{Bmatrix}.
\end{eqnarray}
For the use of remeshed magnetograms as described in Section~\ref{sec:21}, we modify this relation to a discrete representation. To do so, we use the Clenshaw-Curtis quadrature rule given by
\begin{eqnarray}
\int_0^{\pi} d\theta \sin \theta F(\theta) \approx \sum_{i=1}^{N_\theta} \epsilon_i w_i F(\theta_i),
\end{eqnarray}
where $\epsilon_i$ is $1/2$ for $i=0$ or $i=H$, and $1$ elsewhere. The weights $w_i$ are given by
\begin{eqnarray}
w_i = -\frac{2}{H} \sum_{k=0}^{H} \frac{\epsilon'_k}{4 k^2 - 1} \cos \left( \frac{\pi k (i-1)}{H} \right)
\end{eqnarray}
where $H = (H_\theta - 1)/2$, $\epsilon'_k$ is $1/2$ for $i=0$ or $i=H$, and 1 elsewhere. Using this equation, we write Eq.(A9) as 

\begin{align}
\begin{Bmatrix} g_n^m \\ h_n^m \end{Bmatrix} = \frac{2n + 1}{ 4\pi \left(n + 1 + n \left(\frac{R_0}{R_1}\right)^{2n+1}\right)} \frac{2 \pi}{N_\phi} \sum_{i=1}^{N_\theta} \sum_{j=1}^{N_\phi} \epsilon_i w_i P_n^m (\theta_i) B_\text{r}(R_0, \theta_i, \phi_j) \begin{Bmatrix} \cos m \phi_j \\ \sin m \phi_j \end{Bmatrix}.
\end{align}

Using Eq.(A4) and Eq.(A10), we compute the magnetic field components at any point in the region between the solar surface and the source surface~\citep[e.g.,][]{altschuler69,nikolic17}.

\section{SCS Model} \label{app:scs}

The solution for the Laplace equation not bounded by the spherical source surface is given by

 \begin{align}
 \Psi = R_0 \sum_{n=0}^{\infty} \sum_{m=0}^n \left[ \left( \frac{R_0}{r}\right)^{n+1} \left( g_n^m \cos m \phi + h_n^m \sin m \phi \right) P_n^m (\phi)\right].
 \end{align}
 
In the SCS model, solutions for the PFSS model for the magnetic field on the source surface are first oriented to point away from the Sun. This means that if $B_\text{r} < 0$ on the source surface, the sign of all components $B_\text{r}$, $B_\theta$, and $B_\phi$ are reversed. In this study, we match the grid of the PFSS model to the SCS model with equal steps of $2 \degree$ in latitude and longitude. The components of the magnetic field beyond the source surface are

\begin{align}
\begin{split}
B_\text{r} &= -\frac{d \Psi}{dr} = \sum_{n = 1}^{\infty} (n+1) \left( \frac{R_1}{r} \right)^{n+2} {\sum_{m=0}^n{ \left(g_n^m \cos m \phi + h_n^m \sin m \phi \right) P_n^m (\cos \theta )  }}, \\
B_{\theta} &= - \frac{1}{r} \frac{d \Psi}{d\theta} =-  \sum_{n = 1}^{\infty} \left( \frac{R_1}{r} \right)^{n+2} {\sum_{m=0}^n{ \left(g_n^m \cos m \phi  + h_n^m \sin m \phi \right) \frac{dP_n^m (\cos \theta )}{d\theta}}}, \\ 
B_{\phi} &= - \frac{1}{r \sin \theta} \frac{d \Psi}{d\phi}= \sum_{n = 1}^{\infty} \left( \frac{R_1}{r} \right)^{n+2}{\sum_{m=0}^n{ \frac{m}{\sin \theta }\left(g_n^m \sin m \phi  - h_n^m \cos m \phi \right) P_n^m (\cos \theta )}}.
\label{eq:}
\end{split}
\end{align}
In this way, we ensure that no closed magnetic field exists beyond the source surface. We use the least-squares approach to best fit the vector field on the source surface. In order to minimize the sum of squared residuals, we write

\begin{align}
F = \sum_{i=1}^{N_i} \sum_{j=1}^{N_j} \sum_{k=1}^3 \left[ B_k(\theta_i, \phi_j) - \sum_{n=0}^{N_s} \sum_{m=0}^n \left( g_n^m \, \alpha_{nmk} (\theta_i, \phi_j) + h_n^m \, \beta_{nmk} (\theta_i, \phi_j)\right)\right]^2,
\label{eq:}
\end{align}
where $B_k (\theta_i, \phi_j)$ is the reoriented field of the source surface, and the index $k=1,2$, and 3 refers to the radial, latitudinal, and azimuthal fields component of the grid point $(\theta_i, \phi_j)$ of the reoriented magnetic field. Further, $\alpha_{nmk}$ and $\beta_{nmk}$ are

\begin{align}
\begin{split}
\alpha_{nm1} &= (n+1) \cos m\phi \, P_n^m(\theta) \\
\alpha_{nm2} &= - \cos m\phi \, \frac{dP_n^m(\theta)}{d\theta} \\
\alpha_{nm3} &= \frac{m}{\sin \theta} \sin m\phi \, P_n^m(\theta) \\
\beta_{nm1} &= (n+1) \sin m\phi \, P_n^m(\theta) \\
\beta_{nm2} &= - \sin m\phi \, \frac{dP_n^m(\theta)}{d\theta} \\
\beta_{nm3} &= \frac{m}{\sin \theta} \cos m\phi \, P_n^m(\theta)
\end{split}
\end{align}
We compute the derivations to minimize $F$ given by $\partial F / \partial g_n^m$, and $\partial F / \partial h_n^m$. For each $(n,m)$ we can write

\begin{align}
\begin{split}
\sum_i \sum_j \sum_k \alpha_{nmk} (\theta_i, \phi_j) \left( B_k(\theta_i, \phi_j) - \sum_{t=0}^{N_s} \sum_{s=0}^{t} g_t^s \, \alpha_{tsk} (\theta_i, \phi_j) + h_t^s \, \beta_{tsk}\right) = 0, \\
\sum_i \sum_j \sum_k \beta_{nmk} (\theta_i, \phi_j) \left( B_k(\theta_i, \phi_j) - \sum_{t=0}^{N_s} \sum_{s=0}^{t} g_t^s \, \alpha_{tsk} (\theta_i, \phi_j) + h_t^s \, \beta_{tsk}\right) = 0.
\end{split}
\end{align}
The same expression in matrix form is

\begin{align}
\widehat{\alpha \beta} \cdot \widehat B = \widehat{\alpha \beta} \cdot \widehat{\alpha \beta}^\top \cdot \widehat{GH}, 
\end{align}
with 

\begin{align}
\widehat{B} = \begin{bmatrix}
B_1(\theta_1, \phi_1) \\
B_1(\theta_1, \phi_2) \\
\vdots \\
B_1(\theta_I, \phi_J) \\
B_2(\theta_1, \phi_1) \\
\vdots \\
B_3(\theta_I, \phi_J) \\
\end{bmatrix}, \,
\widehat{GH} = \begin{bmatrix}
g_0^0 \\
g_1^0 \\
\vdots \\
g_N^N \\
h_1^1 \\
\vdots \\
h_N^N \\
\end{bmatrix}, \,
\widehat{\alpha \beta}  = \begin{bmatrix}
\alpha_{001} & \hdots & \alpha_{003} \\
\vdots &  & \vdots \\
\alpha_{N_s N_s 1} & \hdots & \alpha_{N_s N_s 3} \\
\beta_{111} & \hdots & \beta_{003} \\
\vdots &  & \vdots \\
\beta_{N_s N_s 1} & \hdots & \beta_{N_s N_s 3} \\
\end{bmatrix}.
\end{align}
Here, the dimension of $\widehat{B}$ is $3 I J \times 1$, the dimension of $\widehat{GH}$ is $(N_s + 1)^2 \times 1$, and the dimension of $\widehat{\alpha \beta}$ is $(N_s + 1)^2 \times 3 I J$. By choosing $\widehat{AB} = \widehat{\alpha \beta} \cdot \widehat{\alpha \beta}^\top$ we find that the solution for $GH$ is of the form 

\begin{align}
GH = AB^{-1} \cdot \alpha \beta \cdot B
\end{align}
This means that the solution of $GH$ requires the inversion of the square matrix $\widehat{AB}$. Finally, we use the estimates of $g_n^m$ and $h_n^m$ from the least mean square fit to compute the magnetic field solution above the source surface. We note that the polarity needs to be refined to match the polarities for $r \le R_1$.

\bibliographystyle{unsrt}  


\end{document}